\documentclass[lettersize,journal]{IEEEtran}
\usepackage{amsmath,amsfonts}
\usepackage{array}
\usepackage{textcomp}
\usepackage{stfloats}
\usepackage{url}
\usepackage{verbatim}
\usepackage{graphicx}
\usepackage{cite}

\usepackage[linesnumbered,ruled,vlined]{algorithm2e}  

\usepackage{subfigure}
\usepackage{array}
\usepackage{multirow}
\usepackage{float}
\usepackage{graphicx}
\usepackage{rotating}
\usepackage{amsmath}
\usepackage{tabularray}
\usepackage{hyperref}
\usepackage{enumerate}
\usepackage{wrapfig}

\usepackage{xcolor}

\begin{document}

\title{C-Koordinator: Interference-aware Management for Large-scale and Co-located Microservice Clusters}


\author{Shengye Song, 
        Minxian Xu,~\IEEEmembership{Senior Member,~IEEE,}
        Zuowei Zhang, 
        Chengxi Gao, ~\IEEEmembership{Member,~IEEE,}
        Fansong Zeng,
        Yu Ding, 
        Kejiang Ye,~\IEEEmembership{Senior Member,~IEEE,}
        Chengzhong Xu,~\IEEEmembership{Fellow,~IEEE}%

\thanks{S. Song, M. Xu, C. Gao and K. Ye are with Shenzhen Institutes of Advanced Technology, Chinese Academy of Sciences, Shenzhen, China. Z. Zhang, F. Zeng, Y. Ding are with Alibaba Group, Beijing, China. C. Xu is with State Key Lab of IOTSC, University of Macau, Macau, China. M. Xu (mx.xu@siat.ac.cn) is the corresponding author.


}
}

\markboth{}%
{Shell \MakeLowercase{\textit{et al.}}: A Sample Article Using IEEEtran.cls for IEEE Journals}


\maketitle

\begin{abstract}
Microservices transform traditional monolithic applications into lightweight, loosely coupled application components and have been widely adopted in many enterprises. Cloud platform infrastructure providers enhance the resource utilization efficiency of microservices systems by co-locating different microservices. However, this approach also introduces resource competition and interference among microservices. Designing interference-aware strategies for large-scale, co-located microservice clusters is crucial for enhancing resource utilization and mitigating competition-induced interference. These challenges are further exacerbated by unreliable metrics, application diversity, and node heterogeneity.

In this paper, we first analyze the characteristics of large-scale and co-located microservices clusters at Alibaba and further discuss why cycle per instruction (CPI) is adopted as a metric for interference measurement in large-scale production clusters, as well as how to achieve accurate prediction of CPI through multi-dimensional metrics. Based on CPI interference prediction and analysis, we also present the design of the C-Koordinator platform, an open-source solution utilized in Alibaba cluster, which incorporates co-location and interference mitigation strategies. The interference prediction models consistently achieve over 90.3\% accuracy, enabling precise 
prediction and rapid mitigation of interference in operational environments. \color{black} As a result, application latency is reduced and stabilized across all percentiles (P50, P90, P99) response time (RT), achieving improvements ranging from 16.7\% to 36.1\% under various system loads compared with state-of-the-art system. These results demonstrate the system’s ability to maintain smooth application performance in co-located environments.\color{black}
\end{abstract}


\section{Introduction}

\begin{figure}
	\centering
	
		\includegraphics[width=0.98\linewidth]{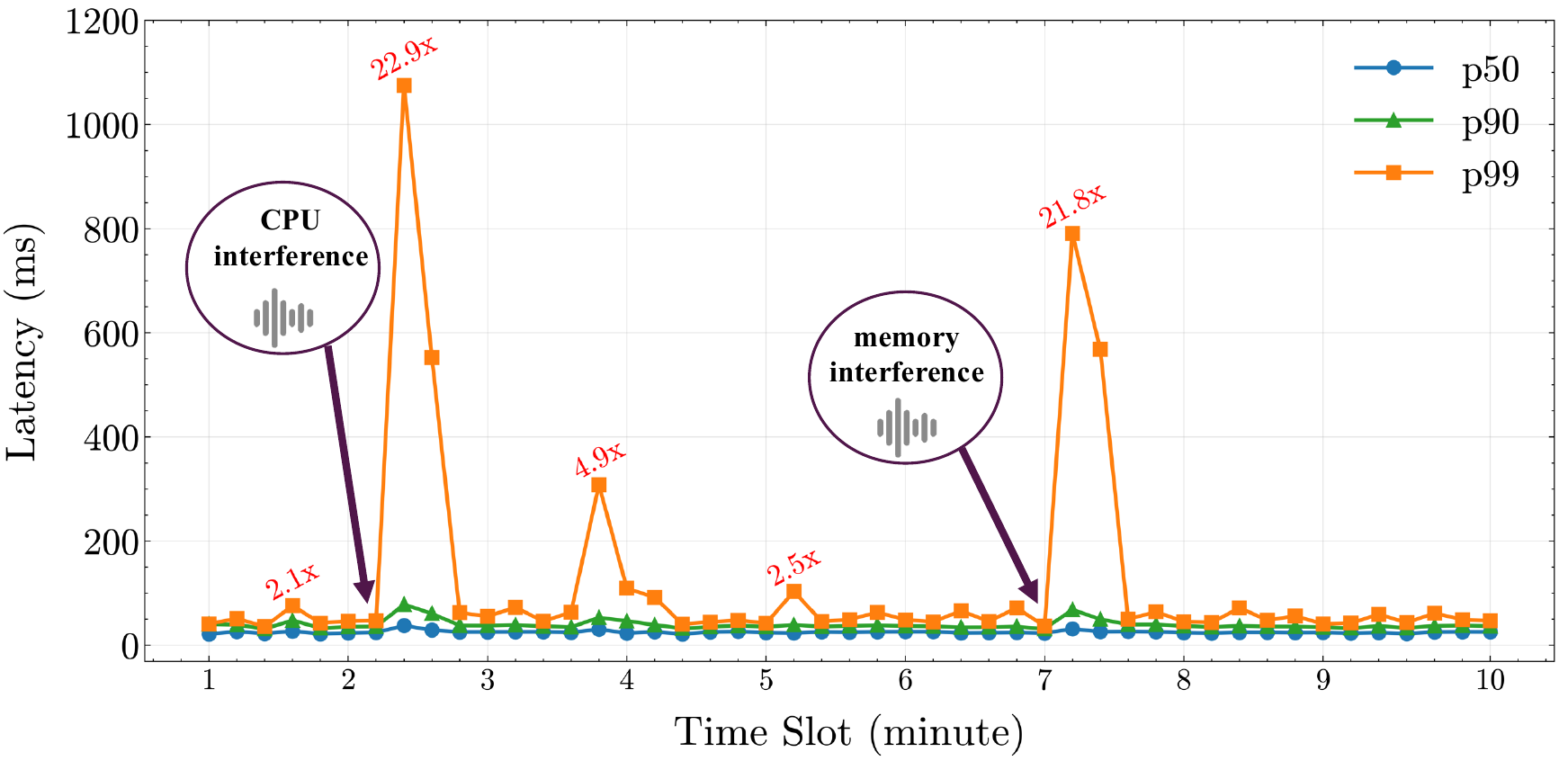}
			
	\caption{Latency and SLO violations under interference in a production Nginx service. At a constant request rate of 100 QPS, the P99 latency remains relatively stable under normal conditions, with occasional fluctuations reaching 2 to 5 times the average baseline. However, when CPU or memory interference is injected, the P99 latency can spike up to 22.9×, leading to severe SLO violations and substantial degradation of user experience.}
\label{fig:interference effect}
\end{figure}

In the evolving landscape of cloud computing, the management of expansive cluster scale presents a unique set of challenges and opportunities. These clusters support a diverse suite of applications, each with distinct functional roles, resource preferences, and core-affinity requirements \cite{zhang2021sinan,delimitrou2014quasar,delimitrou2013paragon}. This diversity, while advantageous, introduces significant complexities in effective cluster management \cite{zhu2022qos,TAAS2025}.
A prominent issue in such a complex environment is the interference scenario where the performance of one application detrimentally affects another. This can manifest in various forms, from increased RT and reduced throughput to significant service latency, directly impairing user experience and potentially resulting in substantial business losses \cite{zhao2021understanding,liu2024suppressing,masouros2020rusty}. To illustrate the real-world impact of such interference in a production environment, we conducted a controlled experiment using a Nginx-based online service deployed in Alibaba’s production cluster. A steady input load of 100 QPS (query per second) was applied to the application, during which we introduced varying levels of CPU and memory interference. As shown in Fig.~\ref{fig:interference effect}, under normal conditions without interference, the P99 latency remained relatively stable, with occasional increased RT by 2.1x-4.9x than the normal. However, once interference was injected, the latency surged dramatically, reaching up to 22.9× of the normal level. This resulted in severe SLO violations and substantial degradation in service quality, which underscores the severe performance degradation caused by interference, highlighting the importance of accurately predicting potential contention and ensuring stable application performance under co-location.

Additionally, interference can cause disproportionate resource allocation, leading to some nodes being overburdened while others remain underutilized \cite{margaritov2019stretch}\cite{CoScal}. Such imbalances degrade overall resource efficiency and increase operational costs significantly \cite{kasture2015rubik}. Furthermore, the interdependent nature of cluster applications, such as microservices \cite{bai2024drpc} with dependencies and share nodes, implies that a failure in one can disrupt others, potentially resulting in partial or total service outages. 

Researchers have been dedicated to enhancing the resource efficiency of data centers through co-location, allowing multiple applications to share the same physical resources \cite{zhao2020large,tirmazi2020borg,verma2015large,delgado2015hawk}. For example, Google’s Borg system \cite{tirmazi2020borg} employs advanced machine learning (ML) models to optimize task allocation and resource utilization. Similarly, Alibaba’s Fuxi system \cite{zhang2014fuxi} orchestrates resource management with precision, ensuring efficient utilization and supporting rapid business iteration. By co-locating multiple applications on the same physical resources, data centers can better utilize idle resources. 

However, while co-location helps to maximize resource utilization, it also exacerbates the issue of interference \cite{patel2020clite,chen2017prophet,chen2017preemptive,iorgulescu2018perfiso}, particularly due to uncontrolled competition for shared resources. This adds another layer of complexity, where not only latency-sensitive applications but also best-effort applications experience performance fluctuations. Furthermore, the inherent dynamic nature of resource utilization within these clusters makes the analysis of applications, nodes, and individual pods even more formidable. For instance, Alibaba's ecosystem \cite{AliwareUnifiedScheduling2021,zhang2022workload} \color{black}comprises millions of nodes, each supporting diverse pods (e.g. can be up to 50 pods) with varying quality of service (QoS) demands (e.g. ranging from millisecond to minutes), resource consumption profiles, and shared resources. Although recent advancements in hardware isolation technologies have shown potential in mitigating interference \cite{wang2017swap,kasture2014ubik,zhang2021libra}, they are largely impractical to apply within Alibaba's infrastructure without considering co-located and large-scale application resource usage characteristics. As a result, interference-induced performance issues are becoming increasingly difficult to detect and predict \cite{zhang2021libra}\color{black}.

In complex ecosystems such as Alibaba Cloud, which are now entirely based on microservices architectures, unhandled interference can significantly impact system performance \cite{guo2024survey,SPE2024}, most notably manifested as increased latency. This latency not only affects the real-time performance of applications but also indicates underlying resource competition that can disrupt the smooth operation of our systems.
This paper outlines our approach within Alibaba’s large-scale, co-located microservices clusters, where we leverage various low-level system metrics to train a predictive model. The model is designed to forecast potential interference by continuously monitoring system performance for anomalies. By detecting these anomalies early, the model enables proactive management of potential disruptions. 
The key \textbf{contributions} of this paper are as follows: 

(1) \textbf{Characterization of Alibaba's applications:}
We characterize the features of Alibaba’s large-scale and co-located applications, underscoring the complexities of detecting interference and identifying robust metrics and methodologies for its effective management. Our analysis of complex metrics aims to seek the optimal metrics and strategies for predicting and mitigating interference, thereby enhancing system resilience and operational efficiency.

(2) \textbf{CPI-based interference prediction model:}
We introduce a CPI-based interference prediction approach to identify the potential interference when applications are co-located. This predictive model achieves over 90.3\% accuracy in forecasting interference for Alibaba's latency critical applications, while also balancing time costs of prediction.

(3) \textbf{Interference-aware management strategy:}
We present Alibaba’s interference-aware management framework, C-Koordinator, to mitigate interference and ensure QoS. \color{black}This framework seamlessly integrates into Alibaba's existing management infrastructure and has been validated for 4 years, enabling efficient and rapid detection of interference, with highly effective interference elimination results.\color{black}

\section{Background and Motivation}
\label{sec:background}
This section highlights the challenges in large-scale and co-located microservice cluster management with Alibaba's scenario, aiming to achieve interference-aware management, and why the existing approaches cannot satisfy our objectives. 

\subsection{Characterization of Alibaba's Microservice Clusters}
We first demonstrate data derived from Alibaba realistic trace to show the co-location related characterization. 

\begin{figure}
	\centering
	\subfigure[PDF shows the Node distribution by number of running applications.]{
		\includegraphics[width=0.23\textwidth]{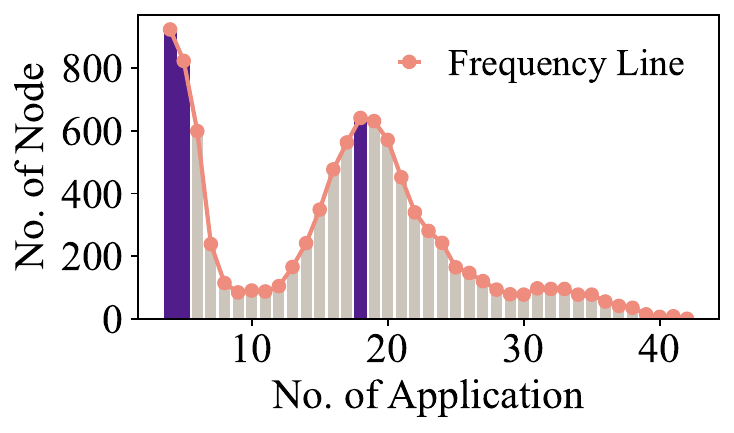}\label{fig:diverse describe(a)}}
			\subfigure[CPU utilization over time for different categories in Node.]{
		\includegraphics[width=0.23\textwidth]{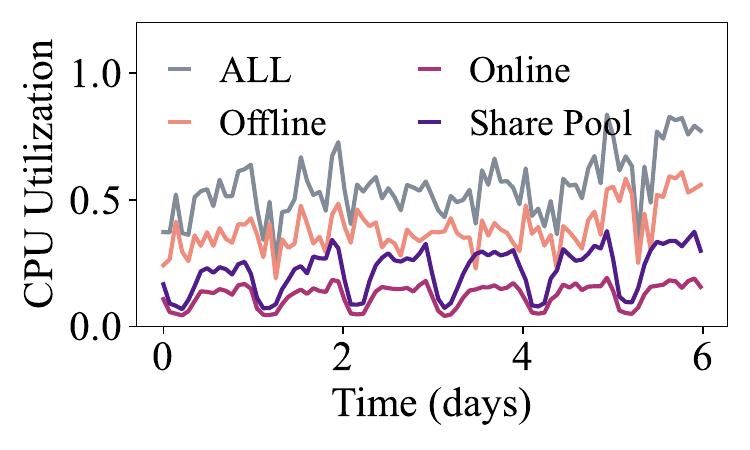}\label{fig:diverse describe(b)}}
	\caption{Co-located application distribution and CPU utilization patterns in Alibaba's cluster.}
\label{fig:diverse describe}
\end{figure}

        
        
            

\textbf{Co-location with Diverse Applications.}  Alibaba’s microservice-based clusters support a variety of applications, such as Taobao (e-commerce), Alipay (financial services), and Alibaba Cloud (infrastructure). As shown in Fig. \ref{fig:diverse describe(a)} that illustrates the complexity and diversity of application distribution within a cluster, which depicts the Probability Density Function (PDF) of the number of applications running on each node, showing a notable variance with approximately 600 nodes hosting around 19 applications each, while about 1000 nodes manage only 4-5 applications. This distribution highlights the intricate and varied nature of node utilization across the cluster.
\color{black}Fig. \ref{fig:diverse describe(b)} presents the CPU utilization over several days for different categories of applications within a single node, including online (stringent QoS requirements for latency and availability), offline (not latency-sensitive and can tolerate being preempted, delayed, or scheduled flexibly), and shared pool (mixed latency requirement) usage. The diagram reveals substantial fluctuations in CPU usage, with high utilization rates that underscore the challenges in resource management and interference prediction within such a dynamic environment. The complexity of these patterns significantly complicates the analysis and prediction of potential resource conflicts and performance interference.\color{black}

The applications in Alibaba's cluster are highly co-located, with over 60\% of nodes running more than 15 applications each. These applications have distinct resource requirements and usage patterns. \color{black}Over 30\% of online service experienced resource contention during peak load periods\color{black}. For instance, during high-traffic events like Singles' Day, e-commerce services can monopolize CPU and memory resources, causing contention and interference with other applications. Additionally, Alibaba’s services demand varied QoS; real-time bidding services in advertising require extremely low latency, while batch processing jobs in data analytics can tolerate higher latency. 
Ensuring these diverse QoS requirements are met without mutual interference is a significant challenge, especially in a co-located environment. Alibaba’s solutions must be exceptionally agile and responsive to handle these spikes without degrading the performance of critical services like Alipay. Existing solutions often struggle to dynamically reallocate resources efficiently across such diverse and fluctuating demands.

\textbf{Complexity in Metrics Collection and Analysis.} Alibaba’s ecosystem generates vast amounts of metrics from user interactions, transaction logs, and system health indicators. Collecting, processing, and analyzing these metrics in real-time to detect interference and performance bottlenecks is challenging due to the sheer scale and diversity of data. Providing real-time performance insights for applications like Alibaba Cloud services requires advanced, scalable monitoring systems. These systems must detect and resolve interference swiftly to maintain QoS across a broad range of applications. 
\color{black}Based on Alibaba's practice, interference should be detected within 1-2 seconds based on performance counters, and should be mitigated within 5-30 seconds. In most cases (over 90\%), interference affecting online services should be fully resolved within 10 seconds, otherwise the users' experience could be significantly degraded.

\begin{figure}[ht!]

        
    \centering
    \includegraphics[width=0.98\linewidth]{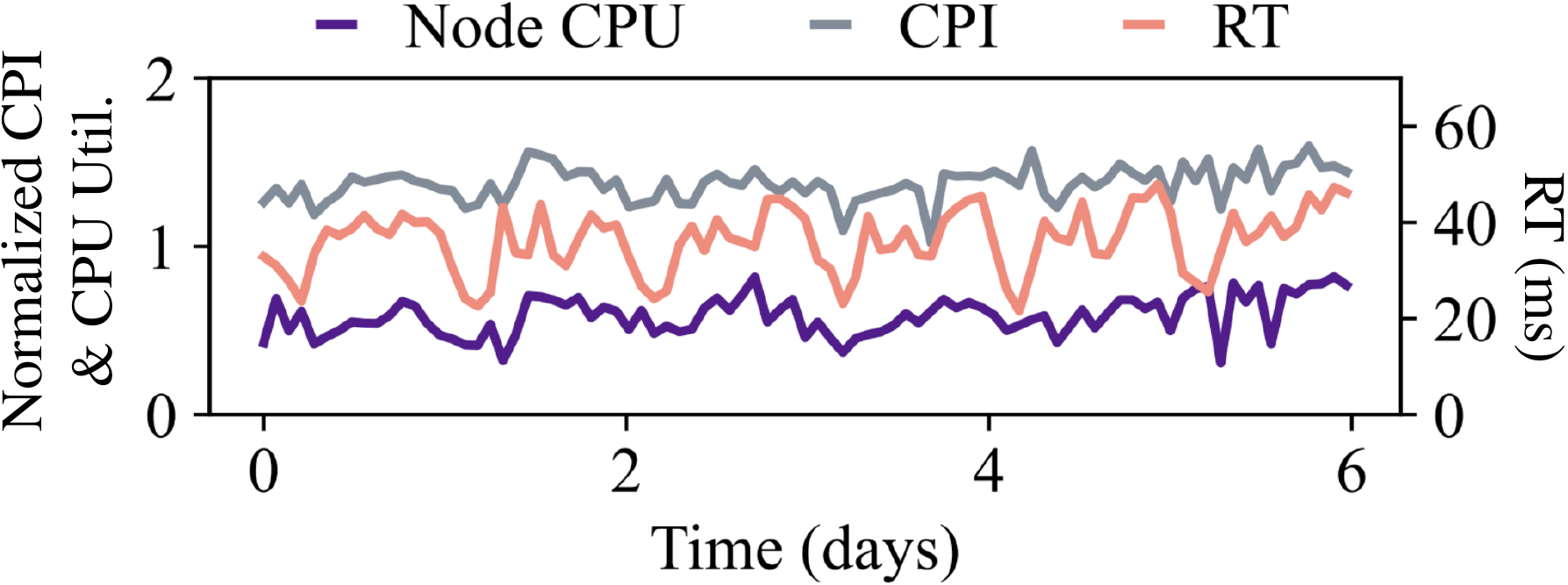}
    \vspace{-0.2cm}
    \caption{Complexity in metrics collection and analysis.}
    \label{fig:metrics}
    \vspace{-0.5cm}
\end{figure}

\color{black}Fig. \ref{fig:metrics} illustrates the intricate relationship between resource utilization and performance metrics within Alibaba's cluster over a six-day period. The diagram shows the fluctuations in node CPU usage, alongside the CPI and \color{black}average RT\color{black}, showing asynchronous and inconsistent changes. The diverse fluctuations across these metrics highlight the dynamic nature of system performance, with notable variations that can be attributed to operational interference. The node CPU utilization curves demonstrate the variability in resource consumption, while the CPI index offers insights into the efficiency of CPU operations under different load conditions. Concurrently, the RT metric provides a direct measure of application responsiveness. This combined view highlights the complexity of correlating these metrics to effectively identify and mitigate performance interference.

\begin{figure}[ht!]
    \centering
    \includegraphics[width=0.984\linewidth]{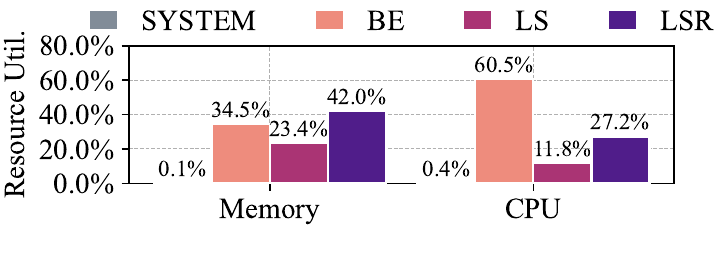}
    \vspace{-0.7cm}
    \caption{CPU and memory usage across application types.}
    \label{fig:usage_distribution}
\end{figure}

The distinct behavior of each metric over the same timeline illustrates the challenges in collecting and analyzing data from multiple sources to diagnose and address potential system inefficiencies, including categories such as \color{black}BE (Best Effort, without strict QoS requirement), LS (Latency Sensitive, with strict latency demand), LSR (Latency Sensitive Reserved, applications with reserved resources), and SYSTEM (operating system services)\color{black}. In addition, as shown in Fig. \ref{fig:usage_distribution}, the high resource utilization across different application types increases the difficulty of maintaining resource isolation at the underlying level. This elevated resource usage raises the likelihood of interference, further complicating system management and performance optimization.

\subsection{Limitation in Metrics Selection}

To address the issue of application performance interference in co-location environments, it is required to timely detect or predict the interference \cite{chen2019co}. One typical approach is to identify interference by monitoring the real-time latency of applications \cite{cui2019tailcutter,delimitrou2016hcloud}, such as employing tail latency to detect interference \cite{lo2015heracles,li2020amoeba}. \color{black}Due to the proprietary management practices of data centers and strict data privacy and security policies, it remains difficult to access fine-grained latency metrics for applications operating in production environments. Moreover, as an application-layer metric, tail latency is inherently influenced by workload characteristics such as QPS, which can fluctuate independently of underlying resource contention. This coupling makes it challenging to isolate and attribute performance degradation directly to interference effects in co-located systems. As a result, researchers often turn to alternative, indirect performance indicators — drawn from both the software system and hardware resource levels — to more accurately assess the presence and severity of interference in multi-tenant environments, like instructions per cycle (IPC) \cite{yang2013bubble}, requests waiting time \cite{lama2018performance}, counter value per instruction (VPI) \cite{pi2022holmes}, system level entropy \cite{liang2023quantifying}, or CPI \cite{verma2015large,zhang2013cpi2}, which become crucial in addressing the interference\color{black}.
However, as the number of resource types to manage increases, the behavior of applications under multi-dimensional resource interference becomes more intricate. Solely relying on indirect indicators to separately evaluate interference in various resource dimensions makes it increasingly difficult to comprehensively assess an application's overall interference susceptibility.

To further refine our metric selection, we conducted observations within Alibaba's cluster, systematically monitoring key performance metrics across different workloads. By analyzing these observations, we identified the metrics that exhibited the strongest correlation with application performance interference. These findings serve as the foundation for the our proposed interference detection model.

\subsection{Limitation in Predictive Model Selection}
\color{black}Given the difficulty of directly measuring interference effects in production cloud environments (e.g. restricted access to fine-grained application metrics and the dynamic, multi-tenant nature of modern data centers), researchers increasingly focus on predictive approaches to model and manage performance interference. Statistical techniques, machine learning models, and deep neural networks, such as decision trees and domain adversarial neural networks, have been employed to construct interference models for applications \cite{javadi2017dial, shi2023alioth}. These predictive models capture the complex relationships between application performance and interference factors across multiple resource dimensions, enabling the estimation of performance variations under diverse workload and co-location scenarios. Compared to direct measurement, predictive models offer a scalable and non-intrusive solution, capable of generalizing to unseen interference patterns and dynamically adapting to workload fluctuations. Once interference is predicted or detected, resource management strategies, including dynamic resource allocation and intelligent job scheduling, can be applied proactively to mitigate performance degradation and ensure QoS.


In the process of constructing interference prediction models, a key step is collecting interference data, with one widely adopted technique being interference-injection-based data collection methods\cite{shi2023alioth,chen2019parties,qiu2020firm}.
This method involves manually introducing interference to analyze applications offline and adjusting resource competition intensity on various shared resources to quantify the degree of application performance degradation. However, given the diversity of applications in large-scale environment, it is infeasible to explore all the cases manually. Currently, the existing prediction models are mainly based on single or limited metrics to predict interference. Additionally, how to balance the prediction accuracy and efficiency in large-scale environment is still an open question. In this work, we aim to provide Alibaba's practice on selection of interference metrics and prediction models\color{black}.

\subsection{Limitations of Existing Solutions}

\textbf{Kubernetes} \cite{burns2022kubernetes} has become the de facto standard for container orchestration, offering a comprehensive and scalable platform for managing microservices. Its key benefits include automated deployment and scaling of containers across clusters. Kubernetes simplifies the management of microservices architectures, providing features such as service discovery, load balancing, storage orchestration, automated rollouts and rollbacks, and self-healing capabilities \cite{carrion2022kubernetes}. 
Additionally, 
the portability and flexibility of Kubernetes make it an attractive choice for organizations seeking to adopt a cloud-native approach and manage their applications consistently across on-premises, hybrid, and multi-cloud environments.

While Kubernetes offers robust orchestration capabilities, it still falls short in interference management due to its lack of built-in mechanisms for handling resource contention. Its reliance on static resource allocation, such as CPU and memory limits, often leads to inefficient resource use and performance degradation among co-located applications. Furthermore, Kubernetes' scheduler lacks the ability to adapt to the dynamic and fluctuating workloads of applications, resulting in resource contention. This issue is particularly problematic in complex, multi-tenant environments where minimizing interference is crucial for maintaining stable and predictable application performance.
Although it integrates basic tools like Prometheus for monitoring, 
the absence of predictive models that forecast resource contention further limits its proactive capabilities, hindering its ability to prevent performance degradation.



\textbf{Koordinator} \cite{Koordinator} cluster management system, an open-source project, aiming at enhancing the Kubernetes's monitoring and predictive capabilities. It is designed to optimize the co-location of microservices, AI, and big data workloads on Kubernetes. As a modern solution tailored for high-demand environments, Koordinator addresses the challenges of resource allocation, utilization, performance management, and interference mitigation in large-scale, diverse workloads. It achieves this through a combination of advanced scheduling techniques and resource management strategies, ensuring that resources are used efficiently while minimizing interference between co-located applications.

Although Koordinator has proven highly effective in large-scale production environments, it exhibits several limitations in interference detection. The system's proactive interference detection capabilities are relatively simple and lack the ability to detect interference with fine granularity and unified way for diverse applications. These limitations result in a coarse detection process that struggled to accurately capture subtle performance fluctuations in complex workloads. 

To address these challenges, Koordinator required a more unified, fine-grained and accurate interference-aware scheduling mechanism. This paper introduces the \textbf{C-Koordinator} (CPI-based Koordinator) system, which builds on the foundation of Koordinator by enhancing its interference detection capabilities. C-Koordinator integrates more granular and precise interference sensing, utilizing advanced monitoring metrics and machine learning (ML) algorithms to predict interference based on CPI. These improvements enable the system to identify potential interference early, allowing for timely interventions that ensure better resource utilization and maintain QoS.

\section{Design of C-Koordinator}
 In this section, we introduce the design of C-Koordinator to address the limitations of the existing approaches and systems. Proven in large-scale production environments within \color{black}Alibaba for more than 4 years\color{black}, C-Koordinator has demonstrated its robustness and effectiveness at scale. Alibaba's diverse range of services, including e-commerce, cloud computing, financial services, and AI tasks, rely on C-Koordinator to maintain high performance and efficiency across millions of containers. 

\subsection{Overall Design Objectives}

The discussions in Section \ref{sec:background} inspire us to achieve the following objectives to achieve interference-aware management for large-scale and co-located clusters:

\begin{figure}
    \centering
    \includegraphics[width=0.99\linewidth]{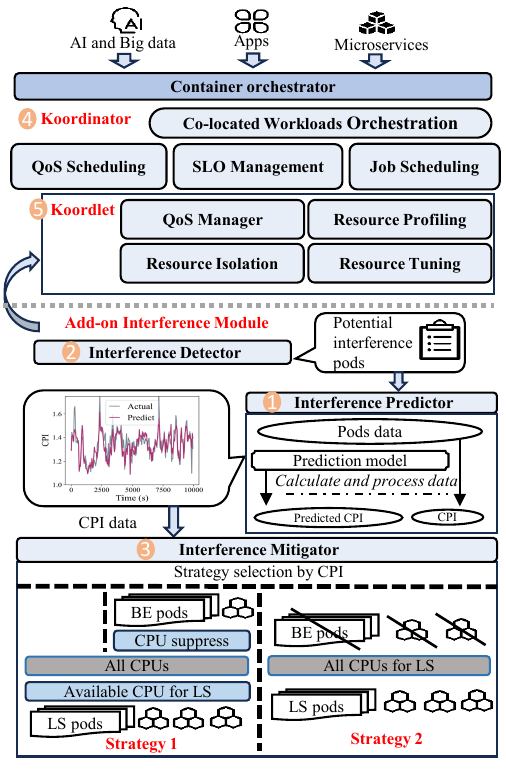}
    \vspace{-0.75cm}
    \caption{C-Koordinator Architecture Overview.}
    \vspace{-0.55cm}
    \label{fig:Architecture}
\end{figure}

\textbf{Unified Interference Detection.} Our system aims to incorporate a unified framework for detecting resource contention and interference across diverse workloads. By leveraging comprehensive metrics such as CPI and other key performance indicators, the system provides a holistic view of interference patterns. This unified detection mechanism ensures that all potential sources of interference are identified promptly, regardless of the application type or resource demands.

\textbf{Proactive Management.} To effectively manage co-located applications, the system must employ proactive resource management strategies. This includes real-time monitoring and predictive analytics to forecast potential interference before they impact application performance. By dynamically adjusting resource allocations based on these predictions, the system can mitigate interference preemptively, ensuring optimal application performance and resource utilization.

\textbf{Fine-grained Monitoring.} The system should offer comprehensive and granular monitoring capabilities that provide real-time insights into resource usage, performance metrics, and interference sources. Advanced monitoring tools and dashboards will facilitate the detailed analysis of resource consumption patterns, enabling quick identification and resolution of performance issues. This fine-grained monitoring ensures that even subtle interference effects are detected and addressed promptly and efficiently.


\color{black}\subsection{Motivation of Using CPI}
In large-scale, co-located microservice-based clusters such as those operated by Alibaba, monitoring and managing performance interference is a persistent challenge due to the dynamic nature of workloads, heterogeneous hardware configurations, and the sheer scale of deployment. Traditional high-level application performance metrics, such as RT and resource utilization, often fall short in this context. RT, while meaningful at the application layer, is influenced by a variety of external factors including QPS, network delays, and upstream/downstream service dependencies, making it difficult to directly attribute variations in RT to underlying resource contention or interference. Moreover, acquiring fine-grained RT data across thousands of production nodes is constrained by privacy, overhead, and access limitations. As our study highlights, the diversity of Alibaba’s application workloads further exacerbates this issue, as different applications exhibit varying latency sensitivities and performance baselines, rendering RT an inconsistent and unreliable interference indicator in large, heterogeneous environments.

To address these challenges, CPI emerges as a more suitable and reliable metric for interference detection and prediction in large-scale, co-located clusters. As a low-level performance counter available on modern processors, CPI reflects the average number of CPU cycles consumed per executed instruction, capturing the combined effects of CPU contention, memory access delays, cache interference, and other micro-architectural stalls. Unlike RT, CPI is largely independent of application-specific request patterns and external service dependencies, providing a uniform, hardware-level signal of performance degradation due to resource interference. Furthermore, CPI can be efficiently collected through lightweight, node-level monitoring tools without exposing sensitive application-layer information, making it scalable and practical for deployment in production-grade clusters. By predicting interference using CPI in conjunction with fine-grained hardware and software metrics across node-, pod-, and application-level data, cloud providers can implement more proactive, precise, and infrastructure-aware resource management strategies to mitigate performance degradation in multi-tenant environments\color{black}.

\subsection{Architecture of C-Koordinator}
As depicted in Fig. \ref{fig:Architecture}, two components are inherited from Koordinator: \textcircled{4}\textbf{Koordinator Scheduler} and \textcircled{5}\textbf{Koordlet} located above the dashed line. The Koordinator Scheduler, deployed as a Kubernetes Deployment, enhances resource scheduling with QoS-aware strategies, differentiated SLO management, load balancing, and resource overcommitment to optimize low-priority workloads. It also manages fine-grained CPU orchestration and QoS policies for memory and bandwidth. The Koordlet manages resource overcommitment, interference detection, and QoS guarantees through modules for CPU, memory, network, and disk profiling, isolation, and contention monitoring. The QoS Manager adjusts node co-location dynamically based on interference detection and SLO configurations, while resource tuning optimizes container performance.

Below the dashed line, in addition to the inherited components, C-Koordinator has significant enhancements to the original Kubernetes and Koordinator, tailored to meet the demands of Alibaba's co-located microservices clusters. By integrating an advanced interference prediction model into the system architecture, the system builds upon core functionalities like QoS Scheduling and SLO Management to deliver a more interference-aware orchestration framework.

The core strength of C-Koordinator lies in its Add-on Interference Module, which integrates three essential components, as highlighted in the diagram: \textcircled{1} \textbf{Interference Predictor}, \textcircled{2} \textbf{Interference Detector} and \textcircled{3} \textbf{Interference Mitigator}. These modules interact seamlessly to manage performance and resource contention at every stage of the application lifecycle. Interference Predictor continuously monitors the system to predict early signs of resource contention. Once interference is detected, the CPI-based Interference Detector identifies potential performance degradation, allowing the system to take preemptive actions.
The Interference Mitigator dynamically adjusts resource allocation to maintain high-priority application performance, especially during peak loads. It works alongside Koordinator’s orchestration capabilities, ensuring efficient interference management while maintaining system stability. The detailed design of these modules is discussed in Section \ref{section:4}. 

\subsection{Design of Key C-Koordinator Modules}

In this subsection, we will discuss the design of the key components in C-Koordinator.

\textbf{Interference Predictor.} Our study in Section \ref{sec:background} demonstrates that common metrics (e.g. RT and resource utilization) cannot accurately reflect the relationship between application performance and inference given the diversity of Alibaba's applications. Therefore, 
We propose an interference-aware approach based on CPI predicted by fine-grained metrics from node-level, pod-level, and application-level for large-scale cluster. 
Traditional metrics often fall short in providing accurate and scalable insights across a vast array of nodes and diverse applications. In our approach, we employ CPI as the primary metric for detecting and predicting application interference. CPI serves as an advantageous metric due to its ability to reflect the performance characteristics of applications across a substantial number of nodes, unlike traditional metrics which may not scale well or accurately represent application performance in heterogeneous environments. This module needs to address which software and hardware metrics should be selected for prediction interference for co-located applications. 

\textbf{Interference Detector.}  In the daily operations of Alibaba's scheduling system, the Interference Detector serves as the fundamental monitoring component. This routine monitoring primarily maintains the QoS for all applications running within the system and monitors the shared resource utilization at each node. Specifically, it tracks the total CPU utilization and total memory utilization  of each node. Additionally, it measures the shared CPU utilization across various applications on node.  If the utilization of these shared resources surpasses a dynamically adjusted threshold, the corresponding application is flagged and added to the list for further interference detection and optimization. This module needs to address which model should be used for predicting interference to balance accuracy and efficiency in large-scale cluster.

\textbf{Interference Mitigator.} After detecting and confirming the presence of interference, the system's interference mitigation policy is activated.  This component is responsible for issuing alerts to the affected application, collecting detailed data on the application's resource usage, and deciding which mitigation strategies to employ. The system dynamically adjusts resource allocations based on the specific characteristics of the affected application and the node on which it resides. This module needs to address which application should be suppressed and which pods should be evicted from the original nodes to reduce resource usage.

\section{Implementation and Practice of the Key Components of C-Koordinator}
\label{section:4}
In this section, we will discuss the implementation details and Alibaba's practice on developing C-Koordinator.

\subsection{Interference Predictor: Fine-Grained Prediction for Interference}

\color{black}CPI is a reliable indicator of application performance, as higher CPI values typically signal inefficiencies caused by resource contention, cache conflicts, or memory bottlenecks. Its hardware-level nature makes it more consistent across diverse applications compared to metrics like RT, which can be influenced by workload patterns, service logic, and external dependencies. However, real-time CPI measurements in production environments are subject to several challenges. CPI naturally fluctuates with changes in instruction set architectures, workload characteristics, system background noise, and transient events like cache misses or branch mispredictions. These fluctuations can introduce considerable noise into real-time data, making it difficult to distinguish between short-lived, benign spikes and sustained interference that requires corrective action. Furthermore, relying on real-time CPI data for interference management introduces operational risks. Instantaneous CPI readings may reflect momentary anomalies rather than meaningful performance trends, potentially triggering unnecessary or suboptimal resource adjustments.

To address this, we propose a predictive CPI-based interference detection approach. Instead of reacting to unstable real-time CPI readings, we use models trained on historical and system-level metrics to forecast CPI trends under varying co-location and workload conditions. This predictive strategy enables proactive interference management, smoothing out short-term fluctuations while capturing meaningful performance trends, thus offering a scalable and accurate solution for interference detection\color{black}.

\textbf{Practice 1: Additional Metrics for Interference Prediction.} In our pursuit to develop an accurate and efficient predictive model for CPI as an interference metric in large-scale and co-located microservice-based clusters, we carefully select the metrics. The challenge lies in balancing the need for comprehensive data to ensure prediction accuracy with the practical constraints of time and resource consumption associated with data collection. Our goal was to filter out a set of key metrics that would provide the most predictive performance while minimizing overhead.

Initially, we focused on the cluster's running status, specifically examining node CPU utilization. So, we introduced commonly used utilization metrics within the container environment, such as node CPU utilization, pod CPU utilization, and memory utilization. These metrics offer a more granular view of resource consumption across different levels of the system. Node CPU utilization provides a broad overview of the total computational load across the entire node, which helps in identifying potential bottlenecks at the node level. However, we recognized that pod CPU utilization is more critical in pinpointing resource usage for individual applications, as each pod represents an isolated environment for specific workloads. By monitoring pod-level CPU usage, we can better assess resource contention and its impact on CPI.
Additionally, memory utilization plays an important role, as memory contention often leads to performance degradation, especially in memory-intensive applications. Monitoring this metric helps in identifying whether resource contention arises from insufficient memory allocation or other factors affecting overall system performance. These metrics were chosen for their ability to provide insights at both the node and pod levels, allowing us to capture the necessary details for accurate interference prediction.

\begin{figure}
	\centering
	\subfigure[Training with util. metrics.]{
		\includegraphics[width=0.23\textwidth]{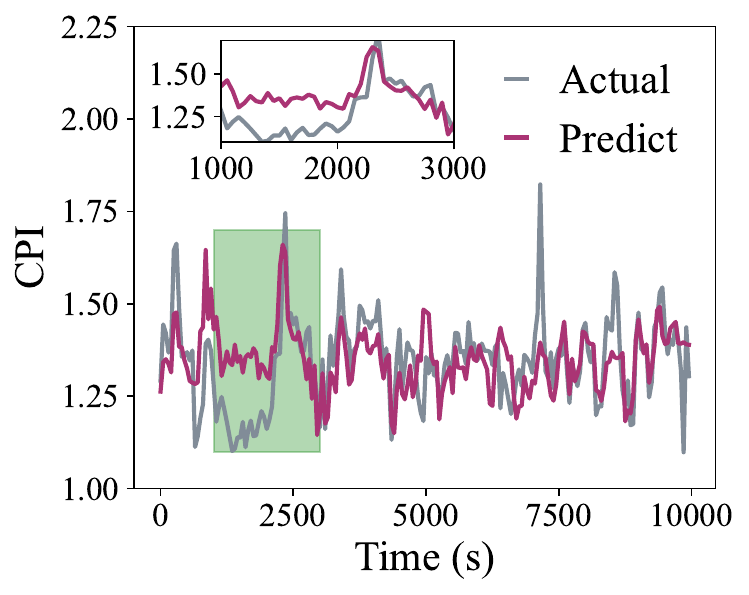}\label{fig:training_comparison(a)}}
			\subfigure[Training with key metrics.]{
		\includegraphics[width=0.23\textwidth]{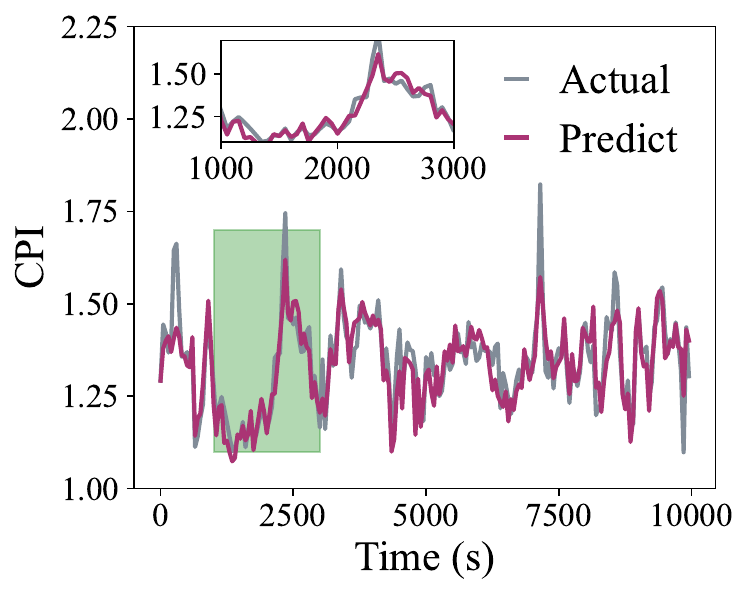}\label{fig:training_comparison(b)}}
	\caption{Training results evaluated across different metrics. (a) Results based solely on resource utilization metrics; (b) Results incorporated the key metrics investigated in this study.}
\label{fig:training_comparison}
\vspace{-0.5cm}
\end{figure}


   
\begin{figure}[t!]
    \centering
    \includegraphics[width=0.99\linewidth]{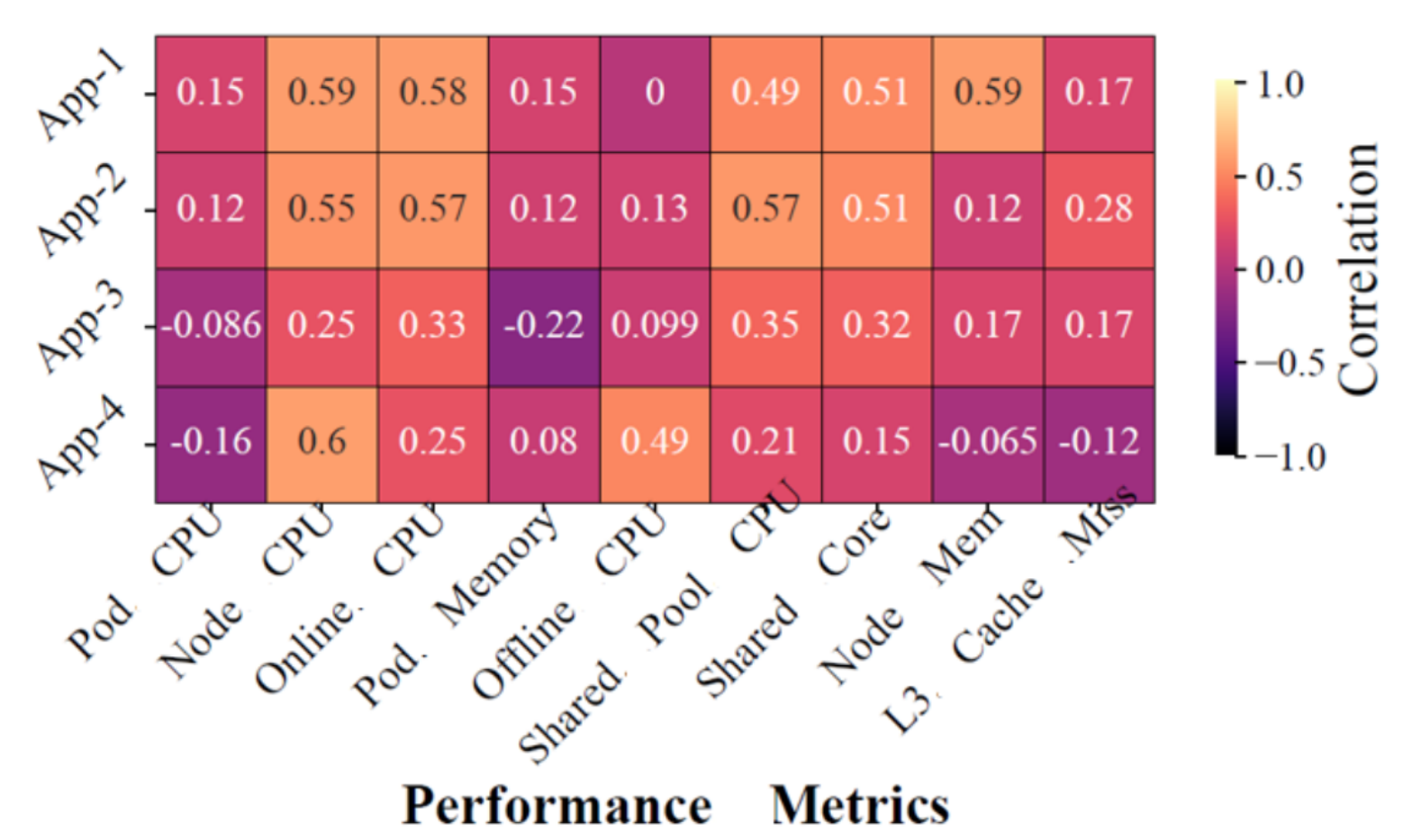}
    \caption{Correlations between performance metrics and CPI across diverse applications. X-axis represents the metrics, and Y-axis displays each application's CPI.
    }
    \label{fig:heatmap}
    \vspace{-0.5cm}
\end{figure}

While these are fundamental metrics, as Fig. \ref{fig:training_comparison(a)} shows that we observed that their effectiveness in predicting CPI is limited in environments where resources are reserved. In such scenarios, actual resource usage tends to be significantly lower than the allocated resources. 
This gap occurs because reserved resources may not reflect real-time consumption, making it challenging to rely solely on these metrics for accurate interference predictions. To overcome this limitation, we expanded our metric set to include more granular indicators that better capture resource contention and performance degradation. As shown in Fig. \ref{fig:heatmap} that the correlation between four different Alibaba's application performance and various low-level metrics varied significantly. This variation highlighted the need for a more tailored approach to interference prediction, as relying solely on general utilization metrics was insufficient. Each application may respond differently to system resource constraints, with some being more sensitive to CPU utilization, while others may be more affected by memory usage or cache performance. 



At the same time, we carefully considered the overhead associated with data collection, such as system resource usage and the time required to gather and process these metrics. After years' observations, we selected a refined set of metrics that not only provided higher accuracy in predicting interference but also minimized the computational costs associated with real-time monitoring. These metrics were chosen based on their ease of collection across all applications and their representativeness of the underlying factors influencing CPI. 

The selected \textbf{nine metrics} include \textit{node CPU }and\textit{ memory utilization, pod CPU }and\textit{ memory usage, core usage (offline and online), shared pool utilization, }and\textit{ L3 cache misses}. We selected core usage as a metric because of its ability to differentiate between resource demands of latency-sensitive online applications and flexible offline workloads. 
In addition,  shared pool utilization is another critical metric, which represents the resources shared by co-located BE, LS, and LSR applications. This resource pool enables dynamic resource sharing based on current application demands, but its complex architecture also makes it susceptible to inefficiencies and contention. If multiple applications with varying resource requirements are simultaneously accessing the shared pool, contention can arise, leading to performance degradation.  
Finally, L3 cache misses were chosen because they signal memory contention among applications sharing CPU cores. Cache misses force the CPU to fetch data from slower memory, causing significant delays. By monitoring these diverse metrics, we gain a comprehensive view of resource contention.  
As shown in Figure \ref{fig:training_comparison(b)}, after a thorough analysis and selection process, the chosen metrics demonstrated a strong ability to predict the CPI, which is highly correlated with application performance. The experimental results confirm that by utilizing these metrics, we can accurately anticipate CPI fluctuations.

\begin{figure}[htbp]  
        \centering
        \includegraphics[width=0.8\linewidth]{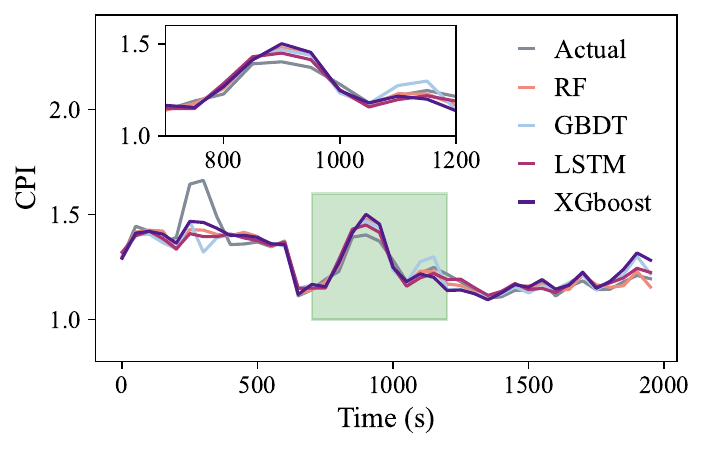}
        \caption{Various models training for same application.}
        \label{fig:third_train}
\end{figure}

\textbf{Practice 2: Selection of Interference Prediction Models.} To identify the most suitable model for Alibaba's cluster environment, we conducted a comparative experiment using real-world data collected from our system. In this experiment, we controlled for the same input metrics (as established in the previous section), ensuring that each model was evaluated on the same dataset. This controlled environment allowed us to focus solely on the performance differences between the models, without any variability from the input data.
We tested a range of models, including traditional ML approaches and more advanced methods like Gradient Boosted Decision Trees (GBDT), XGBoost, Random Forest (RF), Long Short-Term Memory (LSTM), and Multi-Layer Perceptron (MLP).
Our primary focus in model selection was prediction accuracy, and we obtained prediction results as shown in the Fig. \ref{fig:third_train}. The results demonstrated that most models, including XGBoost, MLP, and GBDT, produced predictions that closely aligned with the actual values, making it difficult to distinguish their performance solely based on accuracy.

Given this, we shifted our evaluation towards other critical factors, such as computational efficiency and the ability of each model to meet the demands of Alibaba's large-scale cluster, which is essential for a model to not only deliver accurate predictions but also handle the real-time processing needs of thousands of microservices while maintaining low latency and minimal computational overhead.

As shown in our further analysis in Table. \ref{table:Model Performance}, while models like GBDT, RF, MLP, and LSTM show potential, they lack the efficiency and scalability required for Alibaba’s large-scale environment. GBDT tends to be slower and more memory-intensive, RF struggles with computational efficiency, MLP requires extensive tuning and is less interpretable, and LSTM, designed for sequential data, is too resource-intensive. In contrast, XGBoost combines high accuracy with parallel processing and efficient memory usage, making it ideal for handling Alibaba’s vast data volumes and complex workloads in real-time with minimal overhead. Its ability to handle missing data, optimize memory, and execute gradient boosting efficiently allows it to process large data volumes quickly while maintaining accuracy, which is essential for real-time predictions across thousands of microservices at minimal computational cost. Moreover, its boosting technique iteratively refines predictions by focusing on past errors, achieving high precision without overfitting, even when applied to vast datasets. 


\begin{table}
\centering

\resizebox{\linewidth}{!}{%
\begin{tblr}{
  cells = {c},
  hline{1,7} = {-}{0.08em},
  hline{2} = {-}{},
}
\textbf{Metrics}     & \textbf{XGboost}  & \textbf{RF} & \textbf{GBDT} & \textbf{LSTM}   \\
\textbf{MSE}              & 0.006            & 0.006          & 0.007        & 0.006          \\
\textbf{MAE}               & 0.060            & 0.058           & 0.061        & 0.058          \\
\textbf{R2}                & 0.784            & 0.784           & 0.768        & 0.793          \\
\textbf{ACC}              & 0.948            & 0.947          & 0.938        & \textbf{0.953} \\
\textbf{Training time (s)}       & \textbf{35.745 } & 55.814          & 89.848       & 89.594         
\end{tblr}
}
\vspace{0.5em}  
\caption{Model performance metrics: highlighting best accuracy and training time.}
\label{table:Model Performance}
\vspace{-0.9cm}
\end{table}

\textbf{Practice 3: Working Process of Interference Prediction Model.} Once the system flags an application from the \(List_{Apps}\), which is the output of the Interference Detector module,  list as a potential interference source, it conducts a more in-depth analysis to monitor its performance fluctuations and detect any significant changes. During this detailed monitoring phase, the system collects metrics at multiple levels, including pod-level CPU utilization, L3 cache miss rate ($N_{\text{miss}}$), and pod memory utilization ($M_{\text{pod}}$). These metrics are used to calculate the CPI and assess the overall impact on application performance.

We utilize these metrics as the inputs of a fine-tuned XGBoost-based model. The model processes both historical and real-time data to predict potential performance degradation. Below, we outline the main components and formulas involved in this analysis.

Our model's input metrics are categorized into three groups: 1) Pod-level metric inputs $I_p$ include pod CPU utilization $C_{\text{pod}}$ and pod memory utilization $M_{\text{pod}}$; 2) Node-level metric inputs $I_n$ include node total CPU utilization $C_{\text{total}}^{\text{node}}$, node offline CPU utilization $C_{\text{off}}^{\text{node}}$, node shared pool CPU utilization $C_{\text{sh}}^{\text{node}}$, and node online CPU utilization $C_{\text{on}}^{\text{node}}$; 3) System-level metric inputs $I_s$ include the L3 cache miss rate $N_{\text{miss}}$ and system-wide metrics such as total CPU utilization $C_{\text{total}}$ and total memory utilization $M_{\text{total}}$.

The CPI prediction model utilizes these inputs to estimate potential performance degradation, combining data to calculate
\(CPI_{\text{pred}}\) which is then used to estimate the actual CPI \(CPI_a\) from the pod, node, and system levels through a weighted aggregation of decision trees.
The CPI prediction follows a gradient boosting framework, where the prediction for sample $i$ at iteration $t$ is updated as
\begin{align}
\hat{y}_i^{(t)} = \hat{y}_i^{(t-1)} + \eta f_t(x_i),
\end{align}
with $\eta$ denoting the learning rate and $f_t$ representing the regression tree fitted at iteration $t$. The model optimizes an objective function that combines the training loss and a regularization term, expressed as
\begin{align}
\text{Obj}^{(t)} = \sum_{i=1}^n l(y_i, \hat{y}_i^{(t)}) + \sum_{k=1}^t \Omega(f_k),
\end{align}
where the loss function $l(\cdot)$ is approximated by its second-order Taylor expansion around the previous prediction $\hat{y}_i^{(t-1)}$:
\begin{align}
l(y_i, \hat{y}_i^{(t)}) \approx l(y_i, \hat{y}_i^{(t-1)}) + g_i f_t(x_i) + \frac{1}{2} h_i f_t^2(x_i),
\end{align}
in which the first and second derivatives are
\begin{align}
g_i = \left.\frac{\partial l}{\partial \hat{y}_i}\right|_{\hat{y}_i^{(t-1)}}, \quad
h_i = \left.\frac{\partial^2 l}{\partial \hat{y}_i^2}\right|_{\hat{y}_i^{(t-1)}}.
\end{align}
The regression tree $f_t(x)$ is defined as
\begin{align}
f_t(x) = w_{q(x)},
\end{align}
where $q(x)$ maps input $x$ to a leaf index and $w_j$ denotes the corresponding leaf weight. The regularization term controlling model complexity is formulated as

\begin{align}
\Omega(f) = \tau T + \frac{1}{2} \lambda \sum_{j=1}^T w_j^2,
\end{align}
with $T$ representing the number of leaves in the tree, $\tau$ penalizes tree complexity, and $\lambda$ penalizes large weights. Consequently, the objective simplifies to
\begin{align}
\tilde{\text{Obj}}^{(t)} = \sum_{j=1}^T \left[ G_j w_j + \frac{1}{2} (H_j + \lambda) w_j^2 \right] + \tau T,
\end{align}

where
\begin{align}
G_j = \sum_{i \in I_j} g_i, \quad H_j = \sum_{i \in I_j} h_i,
\end{align}

with $I_j$ being the set of samples assigned to leaf $j$. The optimal leaf weights are obtained by minimizing the objective, given by
\begin{align}
w_j^* = -\frac{G_j}{H_j + \lambda},
\end{align}
and the corresponding minimized objective value is
\begin{align}
\tilde{\text{Obj}}^* = -\frac{1}{2} \sum_{j=1}^T \frac{G_j^2}{H_j + \lambda} + \tau T.
\end{align}
To guide tree growth, the splitting gain is computed as
\begin{align}
\text{Gain} = \frac{1}{2} \left( \frac{G_L^2}{H_L + \lambda} + \frac{G_R^2}{H_R + \lambda} - \frac{(G_L + G_R)^2}{H_L + H_R + \lambda} \right) - \tau,
\end{align}
where $L$ and $R$ refer to the left and right child nodes, respectively. Finally, the overall CPI prediction aggregates the weighted outputs of all trees as: 
\begin{align}
CPI_{\text{pred}} = \sum_{k=1}^K \eta f_k(x) = \sum_{k=1}^K w_k \cdot T_k(I_p, I_n, I_s).
\end{align}
This formulation enables robust integration of pod-, node-, and system-level metrics, facilitating accurate CPI prediction under dynamic cloud computing conditions. Each tree $T_k(\cdot)$ contributes based on its relevance and accuracy, weighted by $w_k$, which is dynamically adjusted in accordance with practical applications at Alibaba.

To enhance the reliability of our predictions, we employ rolling statistics. The rolling mean $RM$, calculated over a predetermined number of samples, provides an average value that helps smooth short-term data volatility. The rolling standard deviation $R_\text{Std}$, on the other hand, measures the variability around this mean, offering insights into the consistency of the data. Their calculation method is as follows:

\begin{align}
RM(x, n) &= \frac{1}{n} \sum_{i=t-n}^{t} x_i,\\
R_\text{Std}(x, n) &= \sqrt{\frac{1}{n} \sum_{i=t-n}^{t} (x_i - RM(x, n))^2}, 
\label{equ:RM&Rstd}
\end{align}
where $x$ represents the time series data, $t$ is the time slot, and $n$ is the rolling window size. For our implementation, \color{black}we set a rolling window of 60 samples with a sampling interval of 5 seconds, capturing 5 minute of data\color{black}. The rolling mean is essentially an average of the data points within the specified window, helping to level out spikes and drops in the CPI collection. The standard deviation of these means further aids in understanding the extent of fluctuation around the average, indicating the stability of node performance.

The difference between the predicted and actual CPI is pivotal in assessing system performance. This discrepancy, \(\Delta CPI\), is used to set a threshold for interference detection. If the variance exceeds a pre-defined limit, it suggests potential issues. 
The system evaluates whether the variance between the predicted CPI (\(CPI_{\text{pred}}\)) and the actual collected CPI data (\(CPI_{Act}\)) surpasses the threshold \(TH_{CPI}\), which is dynamically calculated based on current system conditions. This threshold helps determine whether the fluctuations in performance metrics are within acceptable limits or indicative of underlying issues requiring intervention:


\begin{align}
\Delta CPI &= \left|\text{Mean} \left( CPI_{\text{pred}} - RM\left(CPI_{Act}\right) \right) \right|.
\label{equ:delta_cpi}
\end{align}

To adapt to changing system conditions, the threshold \(TH_{CPI}\) should be more dynamic, considering not only recent fluctuations but also system stress factors like  \color{black}current load \color{black} and resource contention. We introduce a load factor \(L\) for the prediction volatility:
\begin{align} 
T H_{C P I}=k_{1} \cdot R_\text{Std}\left(C P I_{a}\right)+k_{2} \cdot \frac{L_{\text {current }}}{L_{\max }},
\label{equ:equ2}
\end{align}
where $ k_1 $ and $ k_2 $ adjusts the sensitivity of our interference detection, tailored to the operational demands and \({L_{\text {current }}}\) represents the current \color{black} load which is calculated as a weighted sum of normalized CPU, memory, and cache pressure: 
\begin{align} 
L_{\text{current}} = 0.5\times\frac{C_{\text{util}}}{C_{\text{req}}}
                   + 0.3\times\frac{M_{\text{util}}}{M_{\text{req}}}
                   + 0.2\times\frac{N_{\text{miss}}}{N_{\text{max}}},
\label{equ:L_current}
\end{align}
where $C_{\text{util}}$ and $M_{\text{util}}$ are the actual CPU and memory usage of the pod, $C_{\text{req}}$ and $M_{\text{req}}$ are the requested CPU and memory resources, and $N_{\text{miss}}$ is the observed L3 cache miss rate, normalized by the maximum observed value $N_{\text{max}}$ in the cluster. The weights (0.5, 0.3, 0.2) are determined empirically to reflect the relative impact of each resource on CPI and overall pod performance, and can be predefined.

Correspondingly, $L_{\max}$ denotes the maximum possible composite load for the pod  when CPU, memory, and cache pressure all reach their respective normalized upper bounds simultaneously. This normalization ensures that the load factor remains within the range [0, 1], making the detection threshold $TH_{CPI}$ adaptive and comparable across different pods and time windows.

\color{black}
Finally, the system checks whether interference is detected based on the following condition, which is confirmed if the calculated variance \(\Delta CPI \) exceeds the established threshold $TH_{CPI}$: 
\begin{align}
\mathrm{Interference\ Detected} = 
\begin{cases}
1, & \text{if } \Delta CPI > TH_{\text{CPI}} \\
0, & \text{otherwise}
\end{cases}
\label{equ:detect}
\end{align}

when interference is detected, the system logs the affected applications and initiates corresponding actions. The system then calculates the $CSI$ (CPI Severity Index),
\begin{align}
CSI = \frac{\Delta CPI}{TH_{\text{CPI}}},
\label{equ:csi}
\end{align}
This severity index is subsequently passed to the Interference Mitigator module. The Interference Mitigator evaluates $CSI$ and determines the appropriate mitigation strategy.

\subsection{Interference Detector: Timely Identification of Interference}
\label{section:4.2}
Algorithm \ref{alg: interference_alg} shows the combination of Interference Detector, Inteference Predictor and Interference Mitigator. The Interference Detector serves as the fundamental monitoring component. This routine monitoring primarily maintains the QoS for all applications running within the system and monitors the shared resource utilization at each node. Specifically, it tracks the total CPU utilization $C_n^t$ and total memory utilization $M_n^t$ of each node $n$ at time interval $t$. Additionally, it probes the $C_{n}^{s, t}$ (which measures the shared CPU utilization across various applications on node $n$).  

\begin{algorithm}
            \caption{C-Koordinator Interference Algorithm}
            \label{alg: interference_alg}
            \KwIn{$I_p$, $I_n$, $I_s$, $L_{\text{current}}$, $CPI_{Act}$}
            \KwOut{\(List_{Apps}\), $CSI$, Interference mitigation action}
            
            \BlankLine
           \tcc{\textbf{Interference Detector}}\
            \ForEach{node $n$}{
                Monitor $C_n^t$, $M_n^t$, and $C_{n}^{s,t}$\;
                Compute $U_n^t$ using Equation (\ref{equation:U_n^t})\;
                Calculate $TH_{\text{select}}$ using Equation (\ref{equ:TH_select})\;
                \If{$U_n^t > TH_{\text{select}}$}{
                    Flag application and add to \(List_{Apps}\)\;
                }
            }
            
            \BlankLine
            \tcc{\textbf{Interference Predictor}}\
            \ForEach{application in \(List_{Apps}\)}{
                Collect $I_p$, $I_n$, $I_s$, and $CPI_{Act}$\;
                Predict $CPI_{\text{pred}}$ using XGBoost\;
                
                Calculate $RM(CPI_{Act})$ and $R_{\text{Std}}(CPI_{Act})$ using Equation (\ref{equ:RM&Rstd})\;
                Compute $\Delta CPI$ using Equation (\ref{equ:delta_cpi})\;
            }
            
            \ForEach{node in \(List_{Apps}\)}{
                Calculate $TH_{CPI}$ using Equation (\ref{equ:equ2})\;
            }
            
            \ForEach{application in \(List_{Apps}\)}{
                \If{$\Delta CPI > TH_{CPI}$}{
                    Mark as interfered and calculate $CSI$\;
                }
            }
            
            \BlankLine
            \tcc{\textbf{Interference Mitigator}}\
            \ForEach{application with $CSI$}{
                \If{$CSI < \frac{5}{3}$}{
                    Execute strategy 1 in Section \ref{strategy:1}  \;
                }
                \Else{
                    Execute strategy 2 in Section \ref{strategy:2}\;
                }
            }
\end{algorithm}

The comprehensive resource utilization $U_n^t$ of node $n$ at time interval $t$ is computed using the following formula:

\begin{equation}
U_n^t = \alpha \cdot \left( \frac{M_n^t}{1 + M_n^t} \right) + \beta \cdot \sqrt{C_n^t} + \gamma \cdot \left( 2 \cdot C_{n}^{s, t} - \left( C_{n}^{s, t} \right)^2 \right),
\label{equation:U_n^t}
\end{equation}
\color{black}where $\alpha$, $\beta$, and $\gamma$ are weighting coefficients that determine the relative importance of each resource metric—namely, memory utilization, CPU utilization, and shared pool CPU utilization, respectively—in the calculation of node-level comprehensive resource usage. These coefficients are not fixed system-wide, but are instead adaptively assigned for each node to capture node-specific characteristics and workload patterns. For example, on nodes hosting memory-intensive workloads, a higher $\alpha$ value is chosen to emphasize memory pressure, whereas on nodes with high CPU contention, $\beta$ is given greater weight. The value of $\gamma$ can be increased on nodes where the shared CPU pool is a key bottleneck. The coefficients can be determined through historical analysis, profiling, or performance tuning, allowing the model to better reflect resource contention sensitivity unique to each node’s configuration and operational context. This adaptive weighting mechanism improves the accuracy of interference detection by accounting for the heterogeneity and dynamic behavior of nodes within the cluster.\color{black}

Based on our analysis of cluster data, it is evident that resource competition at the node level is primarily influenced by the total utilization of CPU, memory, and shared pool resources on each node. Monitoring at the node level, rather than at the individual pod level, significantly reduces the system's monitoring overhead while ensuring accurate detection of resource contention.

The dynamic utilization threshold $TH_{select}$ is calculated using the following formula:
\begin{equation}
TH_{\text{select}} = \left( \frac{1}{n} \sum_{i=1}^{n} U_{n,i}^t \right) + k \cdot \left( \frac{1}{n} \sum_{i=1}^{n} \left( U_{n,i}^t - \frac{1}{n} \sum_{i=1}^{n} U_{n,i}^t \right)^2 \right),
\label{equ:TH_select}
\end{equation}
where $k$ is an adjustment factor, which is set to 3 in our practice.

If the calculated comprehensive utilization $U_n^t$ exceeds the dynamic threshold \(TH_{\text{select}}\), as $U_n^t > TH_{\text{select}}$,
the system marks the application as a potential source of interference. The flagged application is then added to the \(List_{Apps}\) list for further detailed monitoring, which will be used in Interference Mitigator. Once the comprehensive utilization stabilizes below the threshold for a sustained period, the application is removed from the list.

\subsection{Interference Mitigator: Application Performance Assurance}
\label{section:4.3}
The system computes a dynamic threshold $TH_{CPI}$ for CPI variations and determines whether the interference is mild or severe by comparing the difference $\Delta CPI$ to this threshold.
The severity of interference is categorized based on $\Delta CPI$, for mild interference, when $\Delta CPI$ is less than $\frac{5}{3}$ (this value is configured based on Alibaba's practice) of $TH_{CPI}$ the scenario is treated as mild:

\begin{equation}
CSI < \frac{5}{3} \quad (\text{i.e.,} \, TH_{CPI} < \Delta CPI < \frac{5}{3} \times TH_{CPI}).
\end{equation}
In such cases, the system initiates the CPU suppress strategy, which temporarily reduces the CPU usage of the interfering pods, thereby minimizing its impact on other applications.
For severe interference, indicated when $\Delta CPI$ exceeds $\frac{5}{3}$ of the dynamic threshold: 

\begin{equation}
CSI > \frac{5}{3} \quad (\text{i.e.,} \, \Delta CPI > \frac{5}{3} \times TH_{CPI}).
\end{equation}
The response escalates to the pod eviction strategy, which involves forcibly evicting low-priority pods from the node to free up resources for higher-priority applications and stabilize the system's performance. 

 


After confirming interference, the following strategies will be executed based on the severity of the interference:

\begin{figure*}
	\centering
	\subfigure[CPI prediction for anonymous App-1.]{
		\includegraphics[width=0.32\textwidth]{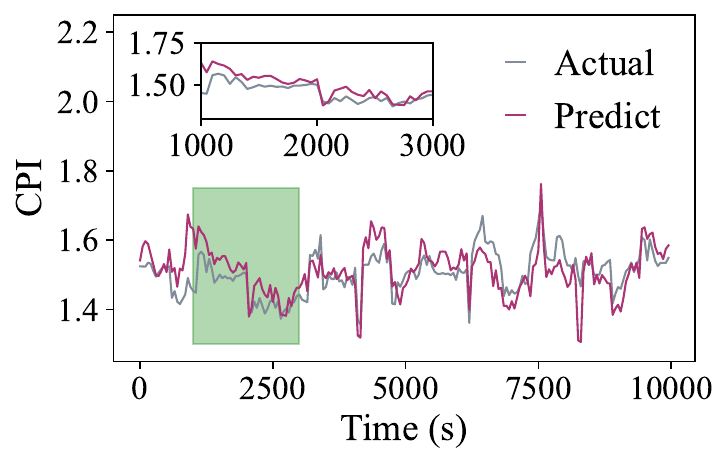}}
			\subfigure[CPI prediction for anonymous App-2.]{
		\includegraphics[width=0.32\textwidth]{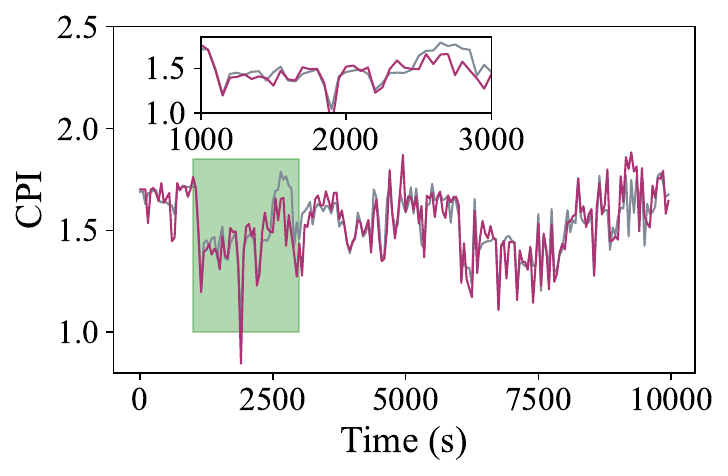}}
			\subfigure[CPI prediction for anonymous App-3.]{
		\includegraphics[width=0.32\textwidth]{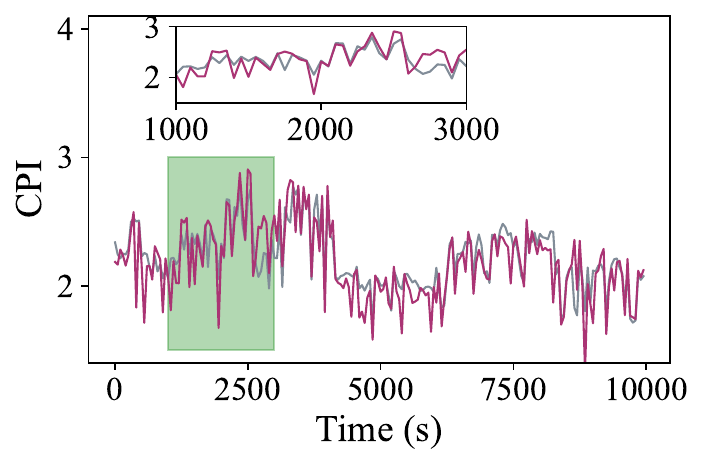}}
	\caption{Evaluating CPI prediction across diverse applications.}
\label{fig:training_APPS}
\end{figure*}

\textbf{Strategy 1: Curtailing CPU Consumption through CPU Suppress.} \label{strategy:1}This strategy is employed in situations where the system experiences mild interference. It works by limiting the CPU usage of low-priority pods, such as BE pods, to ensure that the performance of critical pods, like LS pods, remains unaffected. The CPU Suppress strategy dynamically adjusts the CPU allocation for BE pods based on the current CPU usage of LS pods. The specific CPU limitation is calculated using the following formula:

\begin{equation}
CPU_{\text{restriction}} = CPU_{\text{total}} - (CPU_{LS_{\text{used}}} + CPU_{\text{reserve}}),
\end{equation}
where $CPU_{\text{total}}$ represents the total CPU resources available on the node, $CPU_{LS_{\text{used}}}$ represents the current CPU usage of LS pods, $CPU_{\text{reserve}}$ is a reserved portion of CPU resources allocated to the LS pods to ensure performance stability.

This strategy ensures the performance stability of LS pods by reducing the CPU resources available to BE pods, preventing excessive resource consumption that could degrade overall system performance.

\textbf{Strategy 2: Resource Reclamation through Pod Eviction.} \label{strategy:2}This strategy is applied in cases of severe interference. It directly evicts pods that are consuming excessive resources from the node to rapidly restore node performance.
The pod eviction strategy assesses the CPU usage of BE pods and evaluates their impact on the performance of LS pods. It selects pods for eviction that exceed a predefined CPU usage ratio. The eviction formula is as follows:
\begin{equation}
\text{Evict} = \{ \text{Pod} \mid CPU_{\text{Pod}} > \mu \times CPU_{\text{total}} \},
\end{equation}
where $\mu$ is a predefined threshold for CPU usage, used to determine which BE pods should be evicted, and this value can be configured to control the eviction ratio.

This strategy quickly frees up a significant amount of resources, ensuring that LS pods have sufficient CPU and memory resources to maintain their performance. The combination of CPU suppress and pod eviction allow the system to mitigate the effects of resource contention efficiently, ensuring that critical workloads receive the necessary resources to function optimally.


\section{Performance Evaluations of C-Koordinator}
In this section, we provide evaluations of the C-Koordinator system in terms of its ability to accurately predict and mitigate interference in Alibaba's cluster. 

\subsection{Interference Prediction Evaluation}

In this section, we evaluate the performance of the Interference Prediction Module using real-world data from Alibaba’s internal cluster applications. We selected three representative applications (anonymized for privacy reasons, App-1 represents online web service, App-2 represents database service, and App-3 represents e-business service) that have different CPI fluctuations during runtime to test the module’s accuracy in predicting potential resource contention and interference across a wide range of applications in the cluster.
As demonstrated in Fig. \ref{fig:training_APPS}, our model consistently achieved an accuracy of over 90.3\% for these different applications, showcasing its reliability even with Alibaba’s highly diverse and dynamic workloads. Each application in the cluster exhibits different resource consumption patterns and performance characteristics, yet the prediction module has demonstrated robust and adaptive capabilities, making it highly suitable for large-scale deployment.

However, achieving this level of accuracy does come with certain trade-offs and practical considerations for large-scale cluster. The average training time per application is less than 30 seconds. Importantly, training is only triggered when an application is added to the potential interference list. This ensures that training is conducted only when necessary, keeping overhead minimal.
Based on long-term data from Alibaba’s clusters, about 0.03\% of total applications are regularly added to this list. While this approach does lead to a slight increase in CPU and memory overhead during training, C-Koordinator has already reserved ample resources across all nodes to mitigate resource contention. This pre-emptive resource reservation ensures that system stability remains uncompromised, even when resource usage increases due to interference detection and mitigation.
In summary, despite the slight increase in CPU and memory usage, the C-Koordinator's resource reservation strategy ensures that these overheads remain within acceptable limits, making the module both scalable and efficient for large-scale, real-time deployment.


\subsection{Interference Mitigation Evaluation}


The effectiveness of the Interference Mitigator in C-Koordinator has been evaluated through experiments on a production-scale Kubernetes cluster comprising approximately 7,000 nodes. This cluster hosts a diverse set of workloads, including stateless web services, stateful databases, data processing pipelines, and latency-sensitive real-time applications. The applications are deployed using a variety of resource allocation strategies as well as different CPU binding configurations, including strict CPU pinning, NUMA-aware placement, and CPU share-based scheduling. 

\begin{figure*}
    \centering
    \begin{minipage}{0.99\textwidth}
        \centering
        \subfigure[MySQL P50 latency]{
            \includegraphics[width=0.28\textwidth]{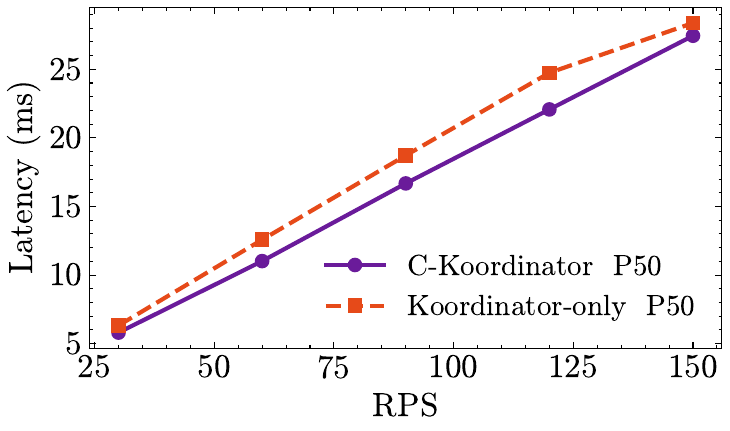}}
        \subfigure[MySQL P90 latency]{
            \includegraphics[width=0.28\textwidth]{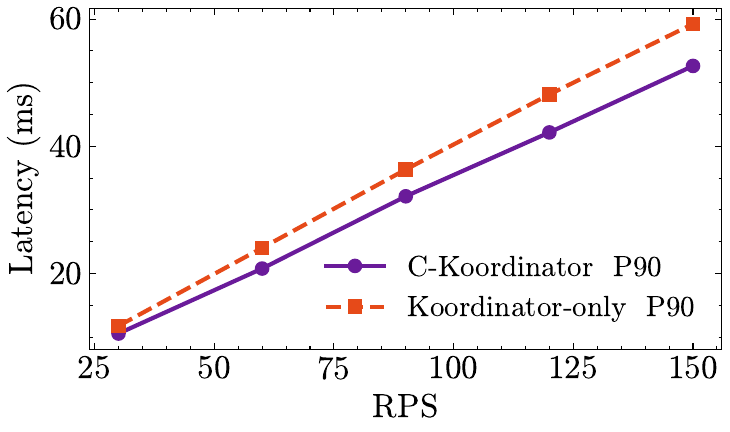}}
        \subfigure[MySQL P99 latency]{
            \includegraphics[width=0.28\textwidth]{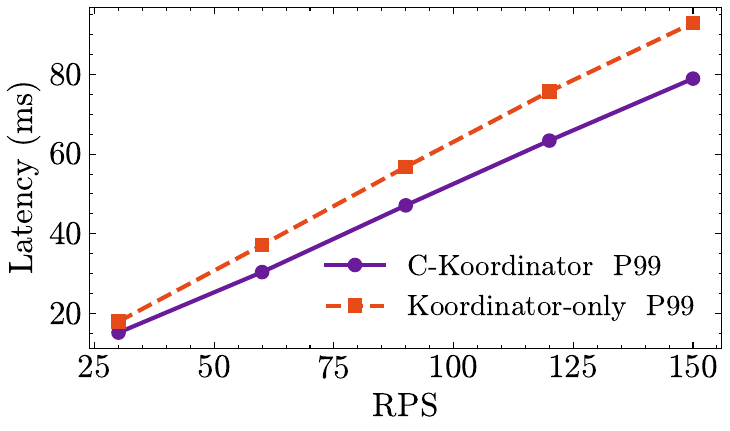}}
        \caption{Latency comparison with MySQL under increasing RPS: (a) P50 latency, (b) P90 latency, (c) P99 latency.}
        \label{fig:pressure1}
    \end{minipage}
    
    \begin{minipage}{0.99\textwidth}
        \centering
        \subfigure[Redis P50 latency]{
            \includegraphics[width=0.28\textwidth]{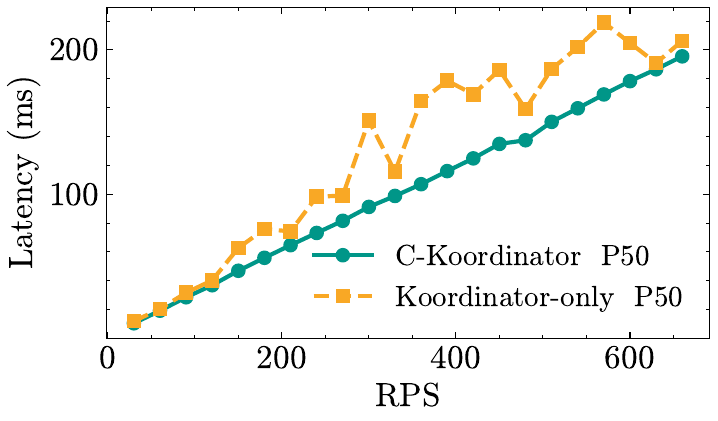}}
        \subfigure[Redis P90 latency]{
            \includegraphics[width=0.28\textwidth]{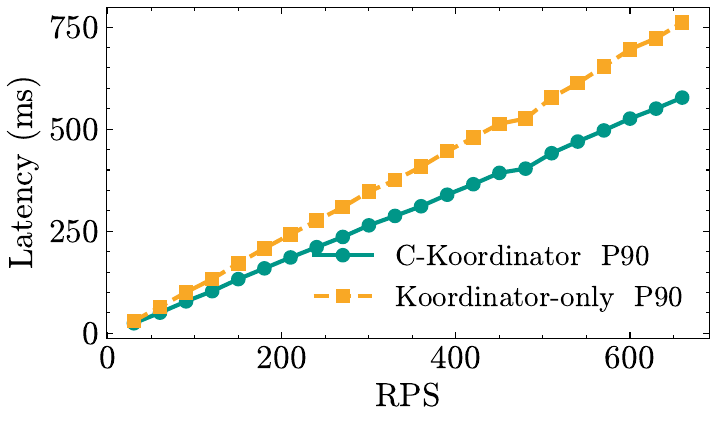}}
        \subfigure[Redis P99 latency]{
            \includegraphics[width=0.28\textwidth]{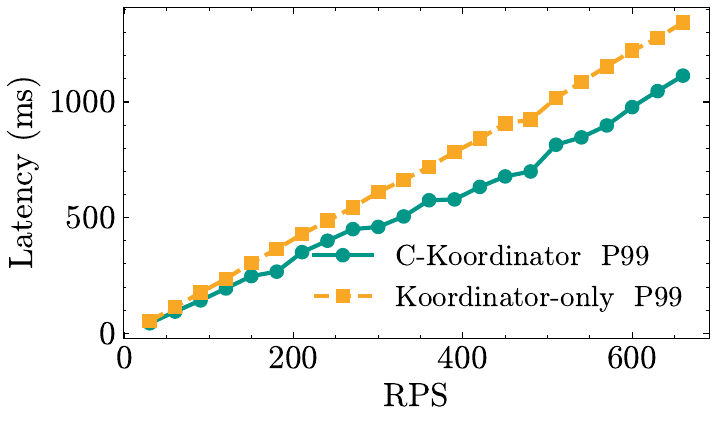}}
        \caption{Latency comparison with Redis under increasing RPS: (a) P50 latency, (b) P90 latency, (c) P99 latency.}
        \label{fig:pressure2}
    \end{minipage}
    
    \begin{minipage}{0.99\textwidth}
        \centering
        \subfigure[Nginx P50 latency]{
            \includegraphics[width=0.28\textwidth]{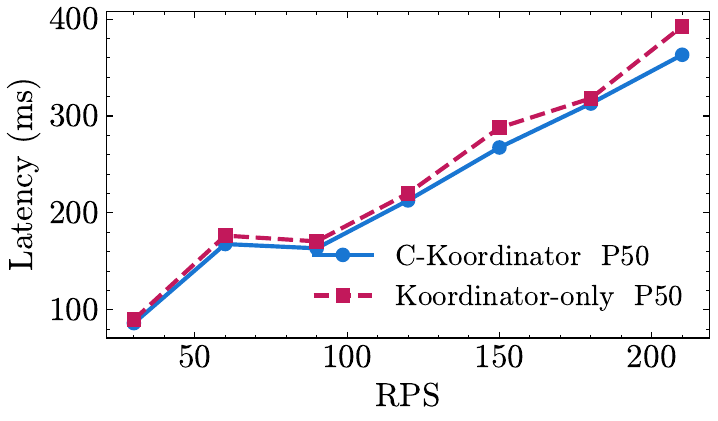}}
        \subfigure[Nginx P90 latency]{
            \includegraphics[width=0.28\textwidth]{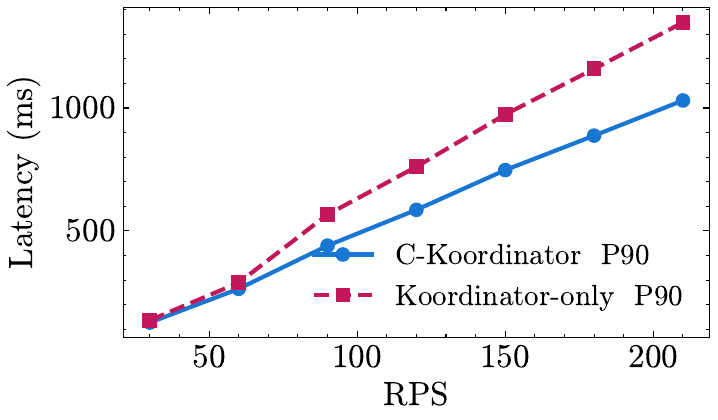}}
        \subfigure[Nginx P99 latency]{
            \includegraphics[width=0.28\textwidth]{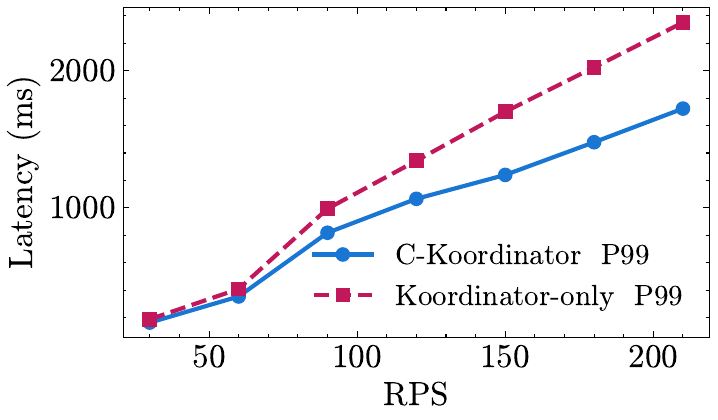}}
        \caption{Latency comparison with Nginx under increasing RPS: (a) P50 latency, (b) P90 latency, (c) P99 latency.}
        \label{fig:pressure3}
    \end{minipage}

\end{figure*}

To make experimental results reproducible, 
we also evaluate the interference mitigation impact on latency reduction under increasing request pressure during normal operation in our 10-node Alibaba g9i instance (4 vCPU and 16 GB memory) cluster with MySQL, Redis and Nginx applications. Compared to the Koordinator, as illustrated in Fig.~\ref{fig:pressure1}-\ref{fig:pressure3}, our approach can reduce latency across all percentiles. For example, for Nginx application as shown in Fig.~\ref{fig:pressure3} that RPS is increased from 0 to 200, latency is reduced by 10.7\% to 36.1\% across various percentiles, especially optimization on P99 tail latency. Similar optimization effects can be observed in Figs.~\ref{fig:pressure1} and \ref{fig:pressure2}, and C-Koordinator can keep P50 latency in Redis increasing more smoothly when RPS is increase significantly compared with Koordinator. The reason results from that C-Koordinator can proactively identify load pressure and reallocate resources in advance, and it can prevent performance degradation and ensures smoother application execution.


\color{black}

The module efficiently managed resource allocation and scheduling, validating its effectiveness in mitigating interference and optimizing performance under different load conditions.
Following the successful reduction in RT, we also assessed the system overhead introduced by the module. This overhead primarily stems from job scheduling processes inherited in Koordinator, triggered during high-pressure conditions when applications are migrated. However, since job scheduling is a native feature of Koordinator, since C-Koordinator already performs numerous scheduling operations daily, this additional overhead is minimal, typically ranging from 1\% to 3\% CPU usage, and remains well within the acceptable limits for the system's operational framework.

\section{Related Work}
In this section, we highlight the relevant work and discuss the limitations of the existing work.

\textbf{Performance Metrics Selection.}
The CPI metric has proven to be an effective indicator of interference for CPU-intensive applications (such as Cpi2 \cite{zhang2013cpi2}) to capture resource contention. Prior work has explored various metrics for interference detection. For instance, Bubble-flux \cite{yang2013bubble} identified a close correlation between the IPC metric and query latency, while Shenango \cite{ousterhout2019shenango} monitored thread and network packet queuing times to gauge CPU resource utilization. PerfCloud \cite{lama2018performance} focused on I/O interference by analyzing variance in blkio wait times, and Holmes \cite{pi2022holmes} introduced the VPI metric to diagnose SMT interference in memory access using hardware performance events. Liang et al. \cite{liang2023quantifying} proposed System-Level Entropy as a method to quantify resource contention through entropy analysis across time series data. Additionally, in co-location environments, competition for shared resources spans multiple dimensions, as highlighted by works like Caladan \cite{fried2020caladan} and Alita \cite{chen2020alita}.

While these methods are insightful, they are limited in their applicability to large-scale clusters. Most of them rely on metrics that are either challenging to collect at scale, such as tail latency \cite{zhao2020rhythm,delimitrou2014quasar,delimitrou2016hcloud}, or depend on detailed hardware monitoring, which may not be feasible in environments with privacy concerns or access restrictions. Moreover, some approaches are focused on specific environments that do not generalize well to the complexity of large-scale systems. For this reason, we utilize CPI, which offers a more scalable and holistic view of application interference, making it more suitable for large-scale and co-located microservice clusters.

\textbf{Interference Prediction Model.}
When applying traditional prediction models to microservice-based environments, the fine granularity and complex chains of microservices demand more sensitive interference detection mechanisms, finer resource isolation measures, and more efficient resource allocation strategies to ensure consistent application performance \cite{luo2022depth,li2022longtale}. While prior research has made significant progress in developing prediction models and orchestration frameworks for such environments, several limitations remain. For example, Lu et al. \cite{lu2023understanding} developed Optum, a unified scheduler for large data centers, focusing on balancing resource utilization and scalability, but it does not fully address the fine-grained interference detection required for diverse, co-located microservices. Similarly, while Luo et al. \cite{luo2022depth} proposed optimization methods for microservice performance in interference-prone environments, their approaches lack real-time adaptability, which is essential for managing the dynamic nature of microservice chains.
Adeppady et al. \cite{adeppady2023reducing} introduced iPlace, a heuristic algorithm designed to minimize deployment interference, but this solution may prove insufficient in large-scale environments where manual adjustments become impractical. Frameworks such as Adrias \cite{masouros2023adrias} and Perph \cite{zhu2021perph}, which leverage deep learning for performance prediction, show promise but encounter overhead and scalability challenges in heterogeneous clusters with diverse workloads. 


However, these models often rely on a limited set of metrics, which may not fully capture the multidimensional nature of interference in real-world, large-scale environments.
In contrast, our work addresses these limitations by providing a more scalable and efficient approach to interference prediction and resource management, specifically designed for Alibaba’s vast microservice architecture. We focus on selecting practical interference metrics and models that strike a balance between prediction accuracy and system efficiency.

\section{Conclusions and Discussions}
In this paper, we have present C-Koordinator that effectively selects relevant metrics and predictive models based on CPI suitable for large-scale and co-located microservice cluster. The accuracy of the predictions has been impressive, with most models achieving over 90.3\% precision, and the selected XGBoost-based model can balance the prediction accuracy and computation efficiency in practice. 
Furthermore, the results have shown significant improvements, demonstrating a noticeable reduction in RT. These findings indicate that the proposed approach is not only feasible but also highly effective in real-world scenarios, ensuring that application performance remains robust even under varying conditions.

Through the development and evaluation of C-Koordinator, several important lessons have emerged that can provide valuable insights for future research and real-world applications. These lessons highlight key aspects of our approach and its impact on system performance and management:

\textbf{CPI as the prediction metric.} CPI is a proactive and effective metric for detecting interference across various applications, capturing performance degradation due to resource contention. Its generalizability makes it ideal for maintaining stable performance in complex multi-tenant systems. 

\textbf{Fine-grained optimization.} Selecting metrics at node, pod, and hardware levels enables precise, fine-grained resource optimization, improving overall system efficiency. This multi-layered approach helps reduce bottlenecks and significantly lowers response latency.

\textbf{Sufficiency of ML models.} ML models are sufficient for accurately predicting CPI, balancing training and inference time effectively. This allows real-time interference detection in large-scale environments without introducing significant computational overhead.

\section{Acknowledgements}
This work is supported by Guangdong Basic and Applied Basic Research Foundation (No. 2024A1515010251, 2023B1515130002), Guangdong Special Support Plan (No. 2021TQ06X990), Shenzhen Science and Technology Program under grants JCYJ20220818101610023, and JCYJ20240813155810014, and Alibaba Air Innovation Research Project.

\bibliographystyle{unsrt1}
\bibliography{sample-base}
\vspace{2cm}
\begin{IEEEbiographynophoto}
{Shengye Song} received his BSc degree from the University of Electronic Science and Technology of China. Currently he is a master student at Southern University of Science and Technology. He conducts scientific research under the guidance of his advisor at the Shenzhen Institutes of Advanced Technology, Chinese Academy of Sciences. His primary research focuses on the interference-aware scheduling in cloud-native systems.
\end{IEEEbiographynophoto}
\vspace{-1.7cm}


\begin{IEEEbiographynophoto}
{Minxian Xu} (Senior Member, IEEE) is currently an Associate Professor at the Shenzhen Institutes of Advanced Technology, Chinese Academy of Sciences. He received his PhD degree from the University of Melbourne in 2019. His research interests include resource management for cloud-native cluster and applications. He has co-authored over 70 peer-reviewed papers published in prominent international journals and conferences with 4600+ citations. He was awarded the 2023 IEEE TCSC Early Career Award (for contributions in efficient management of large-scale microservice-based cluster). 
\end{IEEEbiographynophoto}
\vspace{-1.7cm}

\begin{IEEEbiographynophoto}
{Zuowei Zhang} works at Alibaba Cloud, where he specializes in cloud-native resource optimization and hybrid deployment systems. As a core Maintainer of the Koordinator community, he leads the development of this production-grade, large-scale co-location system, which enhances workload efficiency and cluster utilization while ensuring stability. He has co-authored technical practice on Koordinator's integration with platforms like Kubernetes and YARN, demonstrating significant cost-reduction benefits for enterprises. His work is widely applied in industry-scale cloud environments.
\end{IEEEbiographynophoto}
\vspace{-1.7cm}


\begin{IEEEbiographynophoto}
{Chengxi Gao} (Member, IEEE) received the B.S. and M.S. degrees from the Department of Computer Science, Northeastern University, China, and the Ph.D. degree from the Department of Computer Science, City University of Hong Kong. He is currently an Assistant Professor with the Shenzhen Institutes of Advanced Technology, Chinese Academy of Sciences. He has authored or coauthored more than 20 articles. His research interests include
distributed machine learning and networking systems.
\end{IEEEbiographynophoto}
\vspace{-1.7cm}

\begin{IEEEbiographynophoto}
{Fansong Zeng} is a Senior Technical Expert at Alibaba Cloud and the Contributor of Koordinator, an open-source co-location ecosystem for cloud-native resource optimization. He pioneers large-scale distributed system innovations with over a decade of experience, driving Kubernetes to support Alibaba's core e-commerce systems. His leadership in Koordinator has redefined hybrid deployment paradigms, enabling industry-wide resource efficiency breakthroughs through QoS-aware scheduling and zero-intrusion architecture.
\end{IEEEbiographynophoto}
\vspace{-1.7cm}

\begin{IEEEbiographynophoto}
{Yu Ding} is a researcher at Alibaba, architecting Alibaba's industry-leading 100\% cloud-native migration across millions of cores. As the visionary behind CNCF flagship projects (RocketMQ, Dubbo, KubeVela), his open-source leadership empowers millions of developers. His innovations in hybrid deployment and Serverless computing propelled Alibaba Cloud to become Asia's sole leader in Gartner's container quadrant and Forrester's FaaS Leaders quadrant, setting benchmarks for cloud-native efficiency and scalability
\end{IEEEbiographynophoto}
\vspace{-1.7cm}


\begin{IEEEbiographynophoto}
{Kejiang Ye} (Senior Member, IEEE) received the BSc and PhD degrees from Zhejiang University in 2008 and 2013, respectively. He was also a joint PhD student with the University of Sydney from 2012 to 2013. After graduation, he worked as a postdoctoral researcher at Carnegie Mellon University from 2014 to 2015 and at Wayne State University from 2015 to 2016. He is currently a professor at the Shenzhen Institutes of Advanced Technology, Chinese Academy of Sciences. His research interests focus on the performance, energy, and reliability of cloud computing and network.
\end{IEEEbiographynophoto}
\vspace{-1.7cm}



\begin{IEEEbiographynophoto}
{Chengzhong Xu} (Fellow, IEEE) received the Ph.D. degree in Computer Science and Engineering from the University of Hong Kong in 1993. He is the Dean of the Faculty of Science and Technology and the Interim Director of the Institute of Collaborative Innovation at the University of Macau. He has published two research monographs and more than 300 peer-reviewed papers in journals and conference proceedings. His papers have received more than 20,000 citations with an H-index of 75. His main research interests include parallel and distributed computing, as well as cloud computing.
\end{IEEEbiographynophoto}

\end{document}